\documentstyle[epsf,bibnorm]{lamuphys}
\makeatletter
\let\chapter\hid@chapter
\makeatother
\begin{document}
\pagenumbering{arabic}

\title{Polarization degrees-of-freedom in electronuclear hadron
production from finite nuclei}

\author{J. Ryckebusch 
\thanks{Talk Presented at the Workshop On Electron-Nucleus Scattering,
Elba Internal Physics Center, June 22-26, 1998.},
D. Debruyne and W. Van Nespen} 

\institute{Department of Subatomic - and Radiation Physics, University
of Gent, Proeftuinstraat 86, B-9000 Gent, Belgium}
\titlerunning{Polarization d.o.f. in (e,e$'$N..) reactions}
\maketitle

\begin{abstract}
The polarization degrees-of-freedom in electronuclear two-nucleon
knockout reactions are discussed.  Model calculations for the
unpolarized $^{16}$O(e,e$'$pp) cross sections at ($\mid \vec{q}
\mid$,$\omega$)=(210~MeV,300~MeV/c) are compared to recently obtained
data.  Predictions for the polarization observables in electronuclear
two-nucleon knockout are presented. It is stressed that detailed
studies of two-nucleon emission processes permit to constrain the
contribution from two-nucleon mechanisms (like meson-exchange and
isobar effects) to the single-nucleon knockout channel (e,e$'$p).
\end{abstract}

\section{Introduction}
A systematic study of (e,e$'$p) measurements at x=$\frac {-q^\mu
q_\mu} {2M_N \omega} \approx $1 (quasi-elastic conditions) suggested a
picture of the nucleus that is roughly compatible with {\sl 70\%
mean-field behaviour and 30\% ``correlations''} \cite{vijay}.  In such
a picture, the single-particle spectral function, that determines the
joint probability to remove a nucleon with momentum $\vec{k}$ and to
find the (bound or unbound) residual system at an excitation energy E
can be formally written as \cite{ciofi}
\begin{equation}
P(\vec{k},E)  \equiv  P_0 (\vec{k},E) + P_1 (\vec{k},E) \; ,
\end{equation}
where $P_0$ ($P_1$) is the mean-field (``correlations'') part. The
energy E is usually expressed relative to the ground-state energy of
the target nucleus. The correlations part in the spectral function is
generally conceived as mainly arising from two-body correlations and
its strength is thought to be concentrated around a ridge determined
by the following relation between the energy and momentum
\begin{equation}
\left< E \right> =  \frac {\mid \vec{k} \mid ^2} { 2 M_N} + S_{2N} \;  ,
\end{equation}
where $S_{2N}$ is the threshold for two-nucleon emission out of the
target nucleus.  The above formulae is a formulation of a picture in
which the correlations are assumed to stem from strongly correlated
nucleon pairs with small c.o.m. momentum \cite{strikman}.  With the
aid of the electromagnetic probe, access to the correlated part of the
spectral function is hoped to come from semi-exclusive (e,e$'$p)
measurements that scan high missing-energy regions in the A-1 system
and two-nucleon knockout measurements.  With the latter technique both
``correlated'' partners are detected, yielding accurate information
about the initial conditions of the correlated nucleon pairs.  In this
talk the potency of polarization degrees of freedom in probing the
correlations and (possible) medium modifications in nuclei will be
addressed.  I will concentrate on two-nucleon knockout processes.
Towards the end of the talk, however, it will be stressed that a
better understanding of the two-nucleon degrees-of-freedom (d.o.f)
that results from these two-nucleon knockout studies is a prerequisite
for an unambiguous interpretation of ongoing and planned (e,e$'$p)
measurements.

\section{Structure functions and (polarization) observables in
(e,e$'$NN) and ($\gamma$,NN)}
The cross section for a process in which a reaction of the type  
\begin{equation}
A \; + \overrightarrow{e} (\epsilon) \longrightarrow (A-2)(E_{A-2},\vec{p}_{A-2}) \;
+ N(E_1,\vec{p}_1) \; + N(E_2,\vec{p}_2) \; + e (\epsilon ')
\end{equation}
leads to the excitation of the 
residual nucleus A-2 in a specific state, reads
\begin{eqnarray}
& & {d^8 \sigma \over dE_1 d \Omega _1 d \Omega _2 d \epsilon ' d \Omega
_{\epsilon '}}  (\overrightarrow{e},e'N_1N_2)  =  
{1 \over 4 (2\pi)^8 } p_1 p_2 E_1 E_2 f_{rec}^{-1} \sigma_{M}
\nonumber \\ 
\times & & \Biggl[ v_T {W_T}
+ v_L    {W_L}
+ v_{LT} {W_{LT}}
+ v_{TT} {W_{TT}} 
 + h \biggl[ v'_{LT} {W'_{LT}} + 
v'_{TT} {W'_{TT}} \biggr] \Biggr]
\label{eq:eepnn}
\end{eqnarray}
where $f_{rec}$ is the recoil factor and $\sigma _M$ the Mott cross
section. 
Apart from the dependence on the momentum and energy transfer
(q,$\omega$) from the electrons, all structure functions $W$ exhibit an
explicit dependence on the variables (p$_1$,p$_2$,$\theta
_1$,$\theta_2$ and $\phi _1 - \phi _2$) characterizing the momentum
and spatial direction of the two escaping nucleons.  In addition, the
W$_{LT}$, W$_{TT}$ and W$'_{LT}$ structure functions 
depend on the azimuthal angle of the
center-of-mass $\phi _1 + \phi_2 \over 2$ that can be pulled out of
the structure functions.

\begin{figure}
\centerline{
\epsfysize=6.cm \epsfbox{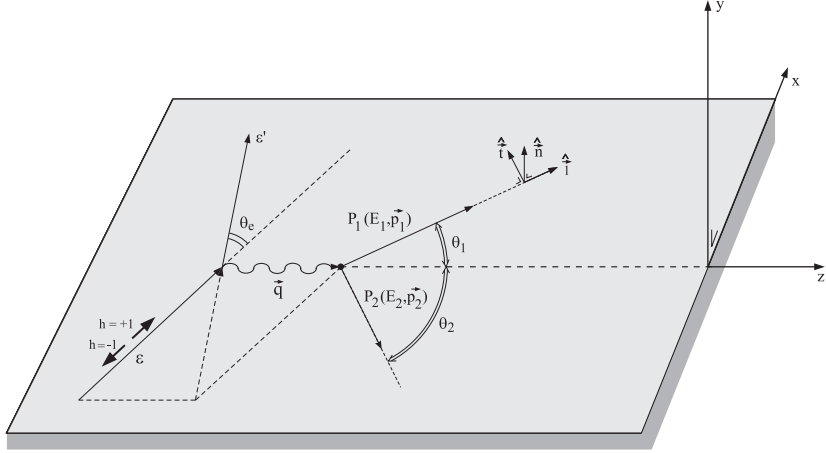}  
}
\caption{\em Reference frame in which the nucleon polarization for the 
($\protect \overrightarrow{e}$,e $' \protect \overrightarrow{N}
\ldots$) reaction is determined. A coplanar situation is considered.}
\label{fig:reffram}
\end{figure}

We remind that apart from a negligible parity-violating component, the
structure function W$'_{LT}$=0 in coplanar kinematics and W$'_{TT}$
vanishes identically, independent of the kinematics.  More favorable
situations are created when performing polarimetry on one of the
ejected hadrons and the spin projection of the latter can be
determined.  In what follows, the polarization of the escaping nucleon
is expressed in the reference frame determined by the unit vectors
(Fig.~\ref{fig:reffram})
\begin{equation}
\hat{\vec{l}} = \frac {\vec{p}_1} {\left| \vec{p}_1 \right|}
\; \; \; \; \;
\hat{\vec{n}} = \frac {\vec{q} \times \vec{p}_1} 
{\left| \vec{q} \times \vec{p}_1 \right|}
\; \; \; \; \;
\hat{\vec{t}} = \hat{\vec{n}} \times \hat{\vec{l}} \; .
\end{equation} 
The escaping nucleon polarization observables can be determined through
measuring {\bf ratios}
\begin{eqnarray*}    
{\mathrm P}_i &=&  \frac {\sigma (s_{1i}\uparrow) -  
\sigma (s_{1i}\downarrow)} 
{\sigma (s_{1i}\uparrow) + \sigma (s_{1i}\downarrow)} \nonumber \\
 {\mathrm P}_i' &=& \frac {
\left[ \sigma (h=1,s_{1i}\uparrow) - \sigma (h=-1,s_{1i}\uparrow) \right] 
- \left[ \sigma (h=1,s_{1i}\downarrow) - 
\sigma (h=-1,s_{1i}\downarrow) \right] }
{
\left[ \sigma (h=1,s_{1i}\uparrow) + \sigma (h=-1,s_{1i}\uparrow) \right] 
+ \left[ \sigma (h=1,s_{1i}\downarrow) +  
\sigma (h=-1,s_{1i}\downarrow) \right] } \;
\end{eqnarray*}
where {$s_{1i}\uparrow$} denotes that hadron ``1'' is spin-polarized in
the positive i direction (i=(n,l,t)) and $h$ is the helicity of the
electron impinging on the target nucleus. 

For the sake of completeness we mention that for real photons, the
unpolarized differential cross section and asymmetry reads in terms of
the structure functions as
\begin{eqnarray}
\frac {d^6 \sigma}
{d \Omega _1 d \Omega _2 dE_1 dE_2}
& = & \frac {1} {(2\pi)^5 2 E_{\gamma}} p_1 p_2 E_1 E_2 
\delta (E_{A-2} + E_1 + E_2 - E_A - E_{\gamma}) \frac {1} {2}
{W_T} 
\nonumber \\
\Sigma & = & 
 \frac {d \sigma _{\parallel} (\vec{\gamma},NN) -
                d \sigma _{\perp} (\vec{\gamma},NN) }
                {d \sigma _{\parallel} (\vec{\gamma},NN) +
                d \sigma _{\perp} (\vec{\gamma},NN) } = 
-  {\frac {W_{TT}} {W_{T}}} \; ,
\label{eq:siggpp}    
\end{eqnarray}
where $\parallel$ ($\perp$) denotes that the photon is polarized
parallel (perpendicular) to the reaction plane determined by the
photon and one of the ejected nucleon's momentum. 

\section{A model for two-nucleon photoproduction on nuclei}
\label{sec:model}
Dealing with both the unpolarized and polarized observables it is
advantageous to calculate the transition matrix elements in the
helicity basis  
\begin{equation}
m_F^{fi} (\lambda = \pm 1,0)
 =   \left< \Psi_{f}^{A-2}(E_x,J_RM_R);
{\vec p}_1 m_{s_{1}}{\vec p}_2 m_{s_{2}}
\mid J_{\lambda =\pm 1,0} ({\vec q}) \mid \Psi _0 \right> \; ,
\label{mfia}
\end{equation}
where $\Psi _0$ is the ground state of the target system, 
$\Psi_{f}^{A-2}(E_x,J_RM_R)$ the (discrete or continuum) state in
which the final nucleus is created and 
(${\vec p}_1 m_{s_{1}}{\vec p}_2 m_{s_{2}}$) the asymptotic momenta and
spin projections (along the z-axis which coincides with the direction
of the momentum transfer) of the ejected particles.  The basic
assumptions of the model that is used to calculate the (e,e$'$NN) and
($\gamma$,NN) cross sections are summarized below.  More details can
be found in Refs.~\cite{jannpa1,janeepp,janprc}.
\begin{itemize}
\setlength{\itemsep}{0.2cm}
\item A consistent shell-model description for the initial and final
state is adopted.  In this manner, orthogonality and anti-symmetry
conditions are naturally obeyed.  Moreover, the calculations for the
A($\gamma$,pp) and A(e,e$'$pp) reaction channels, to which charged
pion exchange is not contributing, are GAUGE INVARIANT.
\item   A distorted wave description for the ejectiles is adopted. 
\item The spectator approximation is adopted.  This implies that only
two nucleons are considered to be directly involved in the reaction
process.  The remaining A-2 nucleons are behaving as spectators.
This assumption does restrict the applicability of the model
calculations to the energy region just above the two-nucleon emission
threshold for which the exclusive character of the
reaction can be guaranteed and there is ample of empirical evidence
that the reaction proceeds in a direct knockout manner.
\item  The cross sections are calculated for each individual state in
the A-2 system.
\item In the model, the center-of-mass and relative motion of the pair
is treated in its full complexity.  We start from a realistic set of
single-particle wave functions obtained from a Hartree-Fock
calculation. With these single-particle wave functions a satisfactory
description of the quasi-elastic (e,e$'$p) data for A$\ge$12 could be
obtained \cite{veerlelarge}.
Unlike for a harmonic-oscillator basis, no formal separation into
relative and c.o.m. coordinates can be pursued.
\item The pionic degrees of freedom are assumed to be the carriers of
the medium-range two-nucleon effects.  We include the equivalent of
all types of Feynman diagrams that are commonly implemented in a
diagrammatic description for pion photoproduction on the nucleon $
\gamma + N \rightarrow N + \pi$.  This includes the Seagull,
pion-in-flight and those diagrams that involve a $\Delta _{33}$
resonance. 
\item One of the major goals of two-nucleon knockout mechanisms is to
enrich our knowledge about the short-range correlations (SRC) and to test
the different models that deal with these effects.  In an attempt to
connect the cross sections directly to the predictions of many-body
theories we start with correlated wave functions of the type
\begin{equation}
 \overline{\Psi} = \frac {\widehat{{\cal G}} \Psi} 
{\left< \Psi \mid \widehat{{\cal G}}^{\dagger}  \widehat{{\cal
G}}  \mid \Psi \right>} \; ,
\label{eq:corwav} 
\end{equation}
where the operator $\widehat{{\cal G}}$ accounts for the corrections
on the Slater determinants  $\mid \Psi >$ 
\begin{equation} 
\widehat{{\cal G}}= {\widehat{\cal S}}
 \prod _{i<j=1} ^{A} \sum _{p} f^p(\vec{r}_{ij},\vec{R}_{ij})
 \widehat{O} ^{[p]} \; ,
\label{eq:corop}  
\end{equation}
where ${\widehat{\cal S}}$ is the symmetrizing operator and $p$
usually runs over a variety of operators.  The correlated wave
function of Eq.~(\ref{eq:corwav}) contains 2,3,$\ldots$,A-body
correlations.  In the calculations, the two-body terms in the
first-order cluster expansion of the transition matrix elements
(\ref{mfia}) are retained.  This approach is justified by remarking
that a recent first-order cluster calculation of the two-nucleon
knockout contribution to the longitudinal $^{12}$C(e,e$'$) strength
illustrated that the two-body terms account for the major effect of
the central short-range correlations \cite{giam}. In our calculations, the
correlation functions $f^p(\vec{r}_{ij})$ are taken from many-body
theories.  The reaction model calculations of which some selective
results will be presented further on should be regarded upon as the
link between the predictions of many-body theories and the actual
data.  It is generally accepted that the (e,e$'$pp) cross sections are
primarily sensitive to the central part in the correlation operator
\begin{equation}
{\widehat {\cal G}} = {\widehat{\cal S}}
 \prod _{i<j=1} ^{A}  f^C(\vec{r}_{ij}) \; \widehat{1} \; ,
\end{equation}
that finds its origin in the hard-core repulsion at short internucleon
distances. 
\end{itemize}

\section{Results and discussion}
\subsection{The $^{16}$O(e,e$'$pp) reaction at low Q$^2$}
\label{sec:nikhef}

\begin{figure}
\centerline{
\epsfysize=14.cm \epsfbox{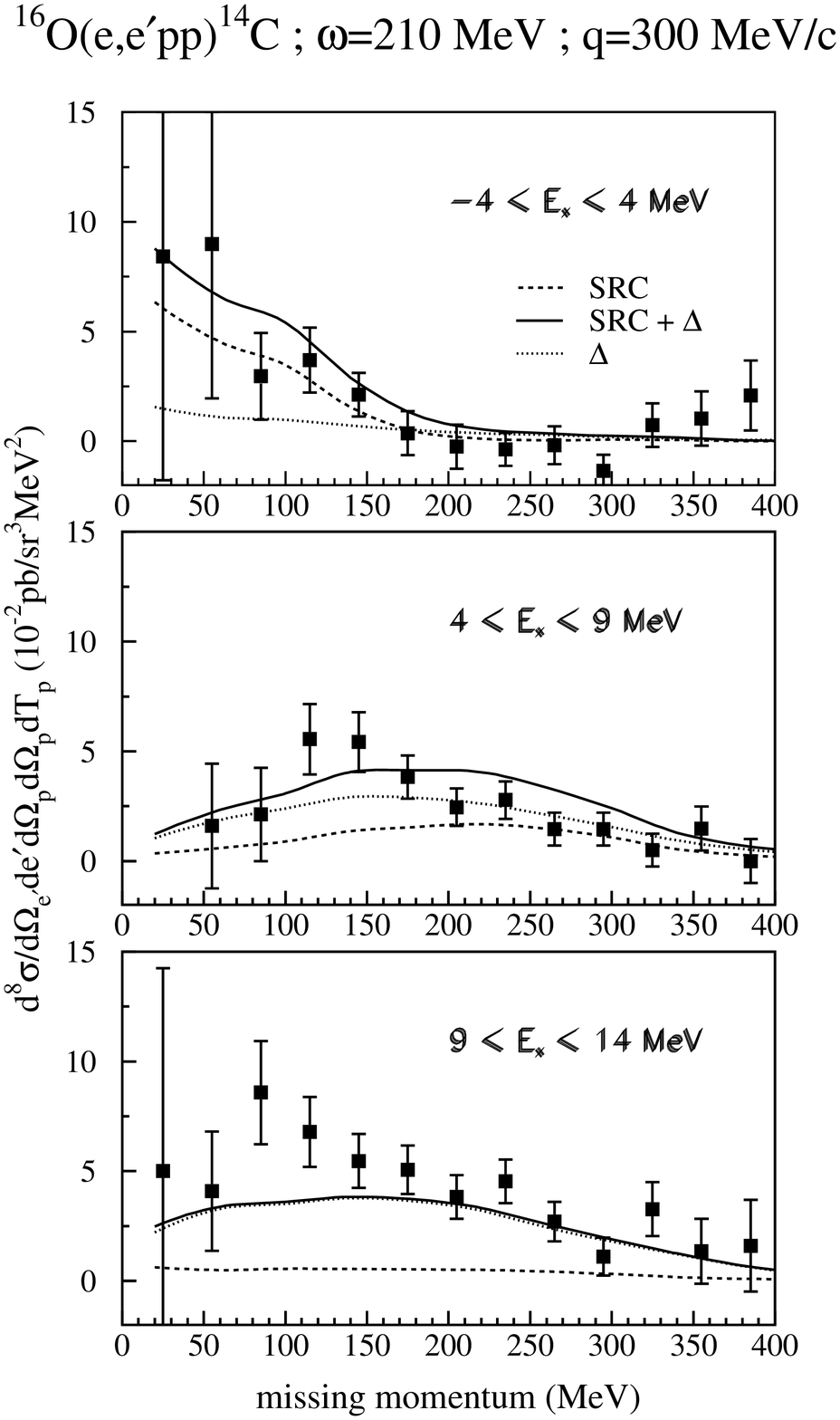}  
}
\caption[ ]{\em Calculated $^{16}$O(e,e$'$pp) missing momentum
distributions for various groups of final states and electron
kinematics determined by e=580 MeV, e$'$=374 MeV and
$\theta_e$=26.2$^o$.  The two-nucleon knockout phase-space covered is
determined by 15$^o \le \theta_1 \le 43^o$, 116$^o \le \theta_2 \le
148^o$ and 52~MeV$ \le $ T$_2 \le $ 108~MeV. The polar angles are
expressed relative to the direction of the momentum transfer. In each
of these variables five mesh points were considered. The data are from
Ref.~\cite{gercoprl2}. The solid line is the result of the
distorted-wave calculation when both the SRC and $\Delta$ isobar
effects are included.  The dotted (dashed) line includes solely the
$\Delta$ isobar (SRC) effects.}
\label{fig:mom}
\end{figure}

Before turning to some predictions for the polarization observables in
two-nucleon knockout studies, a comparison between the model
predictions and some recent $^{16}$O(e,e$'$pp) measurements from
NIKHEF \cite{gerco1,gercoprl2,louk} is presented.  The measurements
utilized hadron detectors that have a rather wide solid angle ($\Delta
\theta \approx 30^o$).  For that reason, the calculations had to be
performed in a grid that covers the full polar and kinetic energy
acceptance of the hadron detectors at a central value for the electron
kinematics ($\omega$=210~MeV/c, q=300~MeV/c).  In the quasi-elastic
(e,e$'$p) case, the dominant peaks in the low-energy part of the
missing energy spectrum can usually be interpreted in terms of one
single-hole component which makes the nuclear-structure input in the
calculations rather simple.  In the two-nucleon knockout case,
however, an additional complication in the calculations stems from the
fact that nuclear-structure calculations \cite{cohen,geurts} point
towards two-body overlap wave functions between the A target and A-2
residual nucleus that can have several two-hole components with a
sizeable amplitude.  The sophisticated nuclear-structure calculations
from Ref.~\cite{geurts}, that were also at the basis of the
theoretical analysis presented in Ref.~\cite{gercoprl2} predict the
following two-proton removal amplitudes for a transition from the
$^{16}$O ground-state to the low-lying states with predominant
two-hole character in $^{14}$C

\begin{eqnarray}
\left| 0^+ ; g.s. \right> & = & 0.77 \left| \left( p_{1/2} \right)^{-2}
; 0^+ \right> + 0.18 \left| \left( p_{3/2} \right)^{-2} ; 0^+
\right> \nonumber \\
\left| 2^+ ; E_{x} = 7.0,8.3~\mathrm{MeV} \right> & = & 
0.11 \left| \left( p_{3/2} \right)^{-2}
; 2^+ \right> 
-0.77 
\left| \left(p_{3/2} \right)^{-1} \left( p_{1/2} \right)^{-1}; 2^+
\right> \nonumber \\
\left| 1^+ ; E_{x}=11.3~\mathrm{MeV} \right> & = & 0.77 \left|
\left( p_{1/2} \right)^{-1} \left( p_{3/2} \right)^{-1}; 1^+
\right> \; .
\label{eq:geurts}
\end{eqnarray}
A striking feature of these removal amplitudes is their smallness.
They are compatible with single-nucleon spectroscopic factors of the
order $\frac {S_{lj}} {(2j+1)} \approx 0.7$ that were systematically
obtained in the analysis of quasi-elastic (e,e$'$p) measurements
\cite{vijay}.  
Indeed, the
two-nucleon spectroscopic factors corresponding with the above removal
amplitudes are
\begin{equation}
\frac {S_{lj,l'j'}} {(2j+1)(2j'+1)}  \approx  0.5 \; ,
\end{equation}
which is approximately equal to the values that one would obtain by
relying on a rather crude estimate based on  
\begin{equation}
\frac {S_{lj,l'j'}} {(2j+1)(2j'+1)} \approx 
\frac {S_{lj}} {(2j+1)} \frac {S_{l'j'}} {(2j'+1)}  \; ,
\end{equation}
and putting $\frac {S_{lj}} {2j+1} \approx$~0.7 as the quasi-elastic
(e,e$'$p) measurements seem to suggest.  Recent $^{16}$O(e,e$'$pp)
measurements at the MAMI facility with superior energy resolution
\cite{guenther1} point towards a strong population of the states at
E$_x$=0.0, 7.0, 8.3 and 11.3 MeV excitation energy in $^{14}$C.  These
are exactly the $^{14}$C states that were observed to be strongly
populated in a $^{15}$N(d,$^3$He) pick-up experiment \cite{dhe3} which
points towards their strong two-hole character relative to the
ground-state of $^{16}$O.  In comparison with the two-nucleon overlap
wave functions as they were derived by Cohen and Kurath \cite{cohen},
the wave functions of Eq.~(\ref{eq:geurts}) are characterized by
smaller amplitudes and a weaker mixing between the $\left| \left(
p_{1/2} \right)^{-2} \right>$ and the $\left| \left( p_{3/2}
\right)^{-2} \right>$ configurations for the ground-state to
ground-state transition.  The results presented below are obtained
with the above wave functions, an exception made for the fact that for
the ground-state to ground-state transition we adopt the relative
mixing between the two configurations as it was predicted by Cohen and
Kurath.  In practice, this amounts to replacing the amplitude 0.18 in
the first line of Eq.~(\ref{eq:geurts}) by 0.31. This operation is
inspired by the observation that with this wave function a more
favorable agreement between the model calculations and the
high-resolution data from Mainz is reached \cite{guenther1}.

The results of the calculations for the three lowest bins in the
excitation energy spectrum obtained from the NIKHEF experiment are
shown in Figure~\ref{fig:mom}.  The comparison between the calculations
and the data is done as a function
of the pair missing momentum 
\begin{equation}
\mid \vec{P} \mid = \mid \vec{p}_1 +
\vec{p}_2 - \vec{q} \mid \; .
\end{equation}
Despite the fact that the experimental resolution in the excitation
(or missing) energy was of the order of 4 MeV and the individual
states could strictly not be resolved, the above considerations allow
to infer that the lowest excitation-energy bin ($-4~MeV \leq E_x \leq
4~MeV$) is as good as exclusively fed through the $^{14}$C
ground-state transition, whereas the second and third bin are mainly
fed through the $\left| 2^+ ; E_{x} = 7.01,8.32~\mathrm{MeV} \right>$
and the $\left| 1^+ ; E_{x}=11.3~\mathrm{MeV} \right>$ states.  The
overall agreement between the calculations and the data is
satisfactory.  In line with the conclusions reached in
Ref.~\cite{carlottub}, a striking feature of the ``SRC'' contribution
is that it is significantly more important for the ground-state
transition in comparison with the other two transitions.  This feature
can be explained as follows.  First, within a harmonic oscillator
basis the sole contributions to the $^{16}$O(0$^+$,g.s.) + e
$\longrightarrow$ e$'$+p+p+$^{14}$C(0$^+$,g.s.) process stems from the
$^1S_0$(T=1) and $^3P_1$(T=1) configurations for the relative diproton
wave function.  As the $^3P_1$ configuration necessarily implies a
c.o.m. P-wave the low missing part of the ground-state transition is a
clear signal of the $^1S_0$(T=1) configuration.  Therefore it is
expected to be very sensitive to the short-range correlations, a
property which is confirmed by the calculations.
Second, it turns out that the $\Delta$ contribution is strongly
suppressed for the $A(0^+) \longrightarrow A-2(0^+)$ transition.  In
an attempt to explain this observation we remark that 
within the context of ($\overrightarrow{\gamma}$,pp) reactions it was
shown that the photon asymmetry $\Sigma$ equals approximately -1 as
long as the initial photoabsorption can be guaranteed to occur on
diprotons residing in a relative $^1S_0$(T=1) state
\cite{sandorfi,janprc}.  In terms of structure functions,
Eq.~(\ref{eq:siggpp}) suggests that under these conditions the $W_T$
will be approximately equal to $W_{TT}$.  At the cross section level
this implies that as long as the initial photoabsorption occurs on a
$^1S_0$(T=1) diproton and the final state interaction has a similar
impact on the $W_{T}$ and $W_{TT}$ structure functions one has
\begin{equation}  
v_T W_T + v_{TT} W_{TT}  \approx  tan^2 \frac{\theta_e}{2} W_T \;
\end{equation} 
suggesting that the $\Delta$ isobar currents, which represent the major
contribution to the transverse channel, will decrease in importance as
smaller electron angles are probed.
A similar sort of cancellation between the $W_T$ and $W_{TT}$ terms,
though an exact one, was noted for the coherent A(0$^+$,g.s.) + e
$\longrightarrow$ A(0$^+$,g.s.)+e$'$+$\pi ^0$ reaction \cite{hirenzaki}.

\subsection{(e,e$'$pp) reactions in super-parallel kinematics}

\begin{center}
\begin{figure}
\begin{minipage}[c]{.35\textwidth}
\caption{\em The missing momentum dependence of the
$^{16}$O(e,e$'$pp)$^{14}$C(0$^+$,g.s.)  differential cross section and
polarization observables in super-parallel kinematics for
typical MAMI kinematics. The solid curve is calculated in the
distorted-wave approximation including the $\Delta$-current and
ground-state correlations.  The latter are implemented through the
central correlation 
function $f^C(r_{12})$ from the G-matrix
calculation of Ref.~\protect \cite{gearhart}. The dot-dashed curve is
the equivalent of the solid line but is calculated with plane wave
outgoing nucleon waves.  The dashed line is the result of a
distorted-wave calculation including only the $\Delta$ current.}
\label{fig:superpar} 
\end{minipage}%
\hspace{0.3cm}
\begin{minipage}[c]{0.55\textwidth}
\centerline{
\epsfysize=15.cm \epsfbox{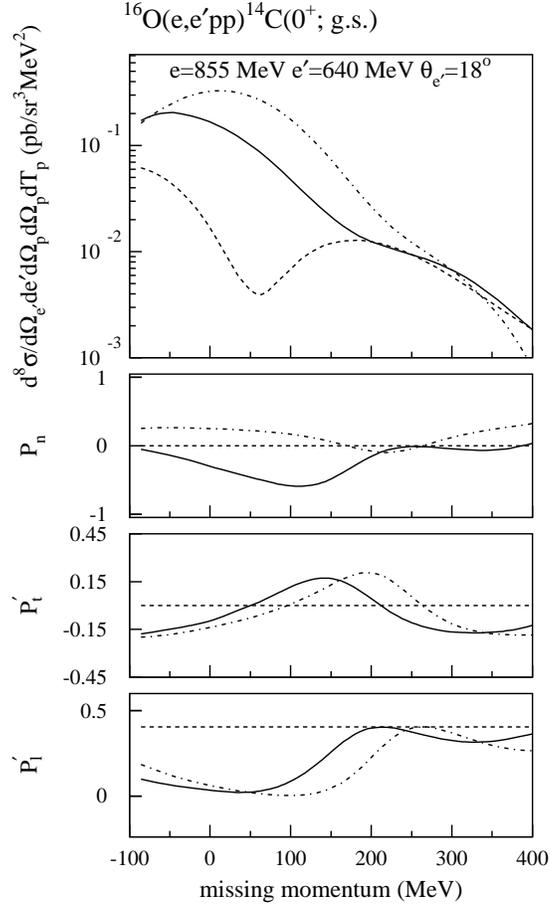}  
}
\end{minipage}
\end{figure}
\end{center}

We now turn to the polarization observables in two-proton knockout and
concentrate on one specific kinematics situation which is judged
favorable for these studies.  So-called super-parallel kinematics
corresponds with the situation that both nucleons are ejected along
the direction of the momentum transfer.  Recent
$^{12}$C(e,e$'$p$\pi^-$) calculations illustrate that resonant pion
production peaks for $\theta _{\pi} \approx 90^o$ \cite{flee}.  For
that reason, $\Delta$ isobar contributions to (e,e$'$pp) that are
originating from initial resonant pion production with subsequent
reabsorption are expected to reach a maximum for both nucleons moving
perpendicular to the direction of the momentum transfer and to be
suppressed in super-parallel kinematics.  Given the large amount of
independent variables that determine the two-nucleon knockout cross
sections, an obvious and probably more interesting asset of
super-parallel kinematics is that only a selected number of structure
functions will contribute to the (e,e$'$pp) cross section and
polarization observables \cite{carlotearly}.  It is to be expected
that this property will facilitate the interpretation of the data.
The selectivity of the cross section and polarization observables to
the different structure functions for planar (e,e$'$pp) processes in
super-parallel kinematics is illustrated in Table~\ref{tab:super}.
Theoretical predictions within the model outlined in
Section~\ref{sec:model} are shown in Figure~\ref{fig:superpar}.  We
have selected the kinematics of the approved MAMI experiment A1/1-97
\cite{a197}.

\begin{center}
\begin{table}
\begin{minipage}[c]{.55\textwidth}
\caption{\em The structure functions determining the cross section (c.s.)
and polarization observables for planar 
(${\protect \overrightarrow{e}}$,e$'{\protect \overrightarrow{p}}$p)
reactions in super-parallel kinematics}
\label{tab:super} 
\end{minipage}%
\hspace{0.6cm}
\begin{minipage}[c]{0.30\textwidth}
\begin{tabular}{|c|c|c|}
\hline
c.s.    & {W$_L$, W$_T$} & (e,e$'$pp)\\
        &              & \\\hline
P$_n$   & {W$_{LT}$}      & (e,e$'\overrightarrow{\mathrm{p}}$p) \\
        &              & \\\hline
P$'_l$  & {W$'_{TT}$}    & ($\overrightarrow{\mathrm{e}}$,
e$'\overrightarrow{\mathrm{p}}$p) \\
        &              & \\\hline
P$'_t$  & {W$'_{LT}$}    &
($\overrightarrow{\mathrm{e}}$,e$'\overrightarrow{\mathrm{p}}$p) \\
& & \\
\hline
\end{tabular}
\end{minipage}
\end{table}
\end{center}

From Figure~\ref{fig:superpar} it is clear that the central
short-range correlations dominate the ground-state transition up to
pair missing momenta of about 150 MeV/c which is in line with the
results obtained in Section~\ref{sec:nikhef} (upper panel of
Figure~\ref{fig:mom}).  Very clear signals of the central short-range
correlations can be deduced from the polarization observables $P_n$
and $P_t'$. Referring to Table~\ref{tab:super} these observables
reflect the interference between the longitudinal and transverse
response.  In the absence of central short-range correlations that
feed the longitudinal channel these observables vanish identically.
The presence of central correlations makes the $P_n$ and $P_t'$
sizeable, particularly for the low missing momentum region where their
contribution is large.  Remark further that the effect of the
final-state interaction is substantially smaller for the double
polarization observables ($P_l'$ and $P_t'$) than for the differential
cross section and $P_n$.

\subsection{Multi-nucleon degrees of freedom and 
($\protect \overrightarrow{e},e' \protect \overrightarrow{p}$)
reactions} 

\begin{figure}
\centerline{
\epsfysize=12.cm \epsfxsize=9.cm \epsfbox{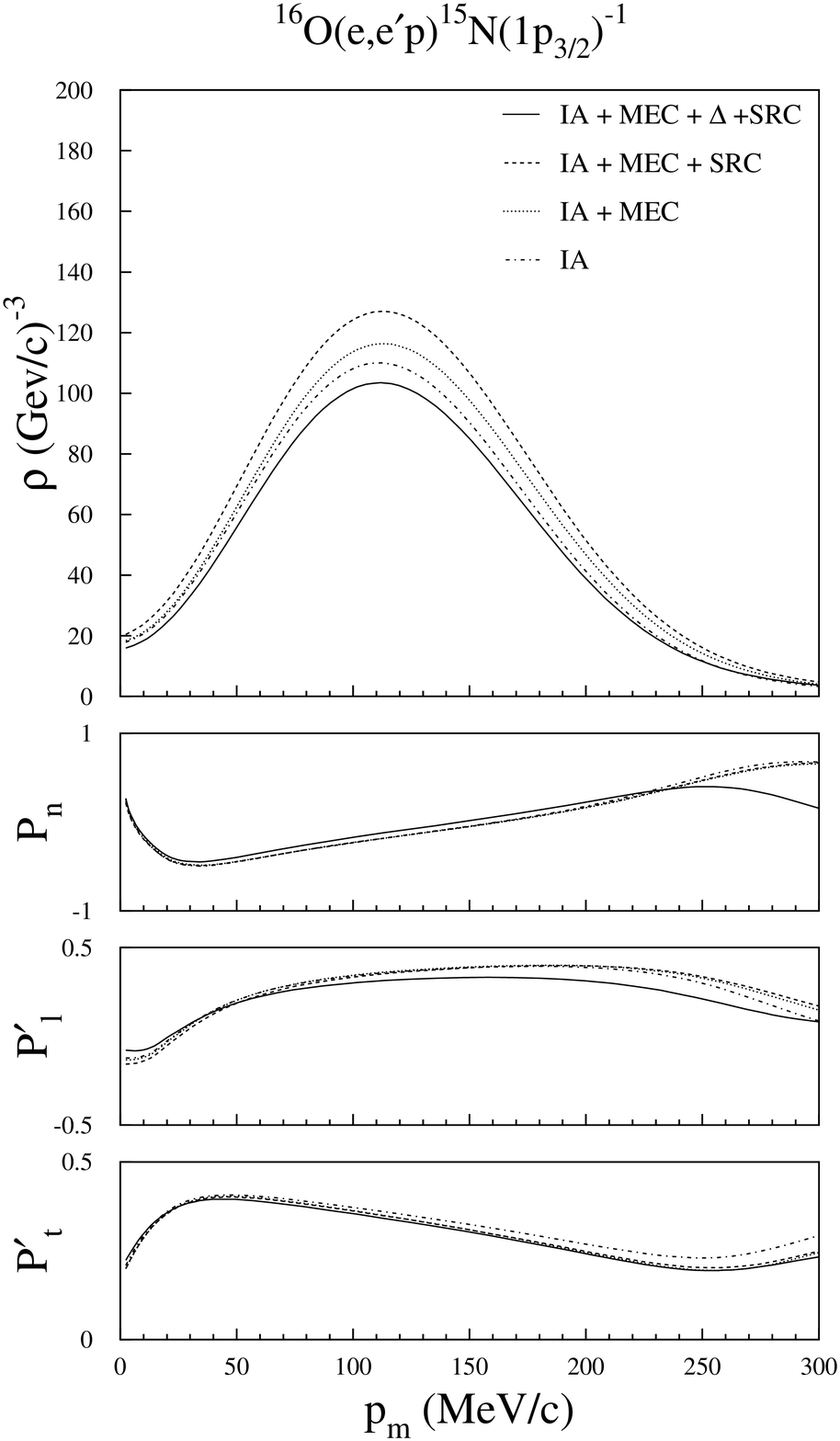}  
}
\caption{\em The sensitivity of the distorted missing momentum
distribution and polarization observables to the various two-nucleon
effects.  Calculations are performed for p$_{3/2}$ knockout from
$^{16}$O at e=850~MeV, e$'$= 610~MeV and
q= 690~MeV/c (Q$^2$=0.42~(GeV/c)$^2$, x=0.938). The variation in
missing momentum is reached by varying the polar angle of the ejected
proton (quasi-perpendicular kinematics).}
\label{fig:dimi1}
\end{figure}

Accumulated information about the two-hadron degrees of
freedom in the nucleus that is gained from two-nucleon knockout
studies with real and virtual photons will be of great value in TJNAF
and MAMI(-C) (e,e$'$p) studies.  Indeed, in many cases A(e,e$'$p)
reactions are meant to provide detailed information about
single-nucleon degrees of freedom and accordingly, meson-exchange and
isobaric currents are unwanted ``background'' that ought be
``theoretically'' controlled.  A few examples of (e,e$'$p)
investigations that fall into this category are

\begin{enumerate} 
\item
The (e,e$'$p) studies at high ($E_m,p_m$), which aim at probing the
``correlated'' part of the spectral function \cite{sick}.  Major
complications in the interpretation of these data in terms of the
single-particle spectral function is the effect of multi-step
processes and multi-body current contributions \cite{louk}.
\item A promising way of testing models that predict substantial
medium modifications \cite{lu98} for the electromagnetic form factors of the
nucleon is high-precision
(${\overrightarrow{\mathrm{e}}}$,e$'{\overrightarrow{\mathrm{p}}}$)
studies at moderate and high Q$^2$ \cite{ron}.  Indeed, in the
plane-wave impulse approximation (PWIA) it can be shown that
\begin{equation}
\frac {P_l'}{P_t'} = - \frac {G_M^p} {G_E^p} 
\frac {(e+e') tan \frac {\theta_e} {2}} {2 M_p} \;  .
\end{equation}
\end{enumerate}

\begin{figure}
\centerline{
\epsfysize=12.cm \epsfxsize=9.cm \epsfbox{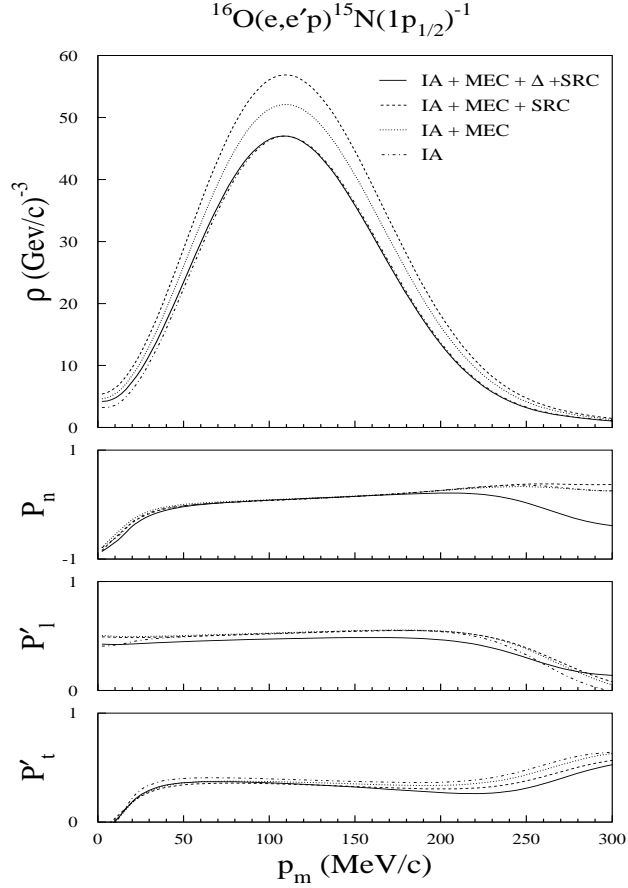}  
}
\caption{\em As in Figure~\protect \ref{fig:dimi1} but for knockout
from the 1p$_{1/2}$ orbit.}
\label{fig:dimi2}
\end{figure}



It is of the utmost importance to investigate all possible mechanisms
that could bring about changes in the above ratio without being
related to (possible) medium modifications of the form factors.
Whereas it was recently shown that final-state interaction and gauge
ambiguities effects are only marginally affecting the ratio $\frac
{P_l'}{P_t'}$ \cite{kelly} the question arises whether meson-exchange
and isobaric currents could bring about any change in the ratio of the
double polarization observables.  In Figures~\ref{fig:dimi1} and
\ref{fig:dimi2} we show the
distorted missing momentum distribution and polarization observables 
$P_n$, $P_l'$ and $P_t'$ for
the $^{16}$O(e,e$'$p) reaction under quasi-elastic conditions. 
Typical kinematics for the MAMI facility
in Mainz was chosen (e=850~MeV, e$'$=610~MeV and
q=690~MeV/c).  No spectroscopic factors were introduced which means that
the calculations are normalized to full subshell occupancy. We have
considered all the two-nucleon effects that are usually included in
the two-nucleon knockout calculations : meson-exchange currents (MEC),
$\Delta$-isobar currents (IC) and the effect of central short-range
correlations.  With respect to the short-range effects we want to
stress that in Ref.~\cite{janeepp} it was shown that in the lowest
order cluster expansion the short-range effects as they
are introduced through equation Eq.~(\ref{eq:corwav}) can be
implemented by considering a two-body operator of the type
\begin{eqnarray}
-\sum _{i<j} & & \left[ \left(\vec{J}^{[1]} (i) + \vec{J}^{[1]} (j)
\right) g(\vec{r}_{ij}) + g^{\dagger} (\vec{r}_{ij})
\left(\vec{J}^{[1]} (i) + \vec{J}^{[1]} (j)
\right) \right.
\nonumber \\
& & + \left. \vec{J}^{[2]} (i,j) g(\vec{r}_{ij})
+ g^{\dagger} (\vec{r}_{ij}) \vec{J}^{[2]} (i,j) \right] 
\end{eqnarray}
where $\vec{J}^{[1]}$ is the one-body current operator as it would be
considered in the impulse approximation, $\vec{J}^{[2]}$ the two-body
current operator including the meson-exchange (MEC) and
$\Delta$-isobar (IC) currents and $g(\vec{r}_{ij})$ is a shorthand
notation for $1-f^C(\vec{r}_{ij})$.  The contribution from the
two-nucleon currents to the single-proton knockout channel was
calculated by explicitly summing over all occupied proton and neutron
single-particle states in the target nucleus \cite{veerlemec}.
Referring to Figures~\ref{fig:dimi1} and \ref{fig:dimi2}, a striking
feature is that the two-nucleon effects do not significantly alter the
shape of the effective missing-momentum distributions for the low
missing-momentum region.  As a consequence, their effect would not be
noticed when comparing IA calculations with data but would simply be
``effectively'' accounted for in the spectroscopic factor that is
introduced to scale the calculations to the data.  The net effect of
the various two-nucleon effects is a reduction of the $1p_{3/2}$ cross
section and almost negligible in the cross section for its spin-orbit
partner $1p_{1/2}$.  In line with the findings of
Ref.~\cite{veerlemec} the impact of the two-nucleon effects is
generally bigger for the $j=l+\frac{1}{2}$ than for the corresponding
$j=l-\frac{1}{2}$ single-particle state.

The effect of the two-nucleon currents on the polarization observables
is rather small in the low missing momentum region for the considered
quasi-elastic kinematic conditions (x=0.94).  On the other hand,
particularly the $P_l'$ observable is predicted to exhibit some
sensitivity to the $\Delta$-isobar currents, an effect which warrants
further investigation in the light of previous discussions.  With
increasing $Q^2$ the relative importance of two-body currents is expected to
decrease.  In this regime, however, numerical calculations with
two-body currents become very involving as a large number of
multipoles in the expansions of the electromagnetic current operators
is required before convergence can be reached.
   
\section{Conclusions and Outlook}
In conclusion, there is accumulating evidence that both scalar
short-range correlations and $\Delta$ isobar effects from $\gamma^* pp
\longrightarrow \Delta ^+ p \longrightarrow p p \pi^0 \longrightarrow
pp $ contribute to the A(e,e$'$pp) reaction.  Real photon studies with
polarized photons will fine-tune the different $\Delta$ and MEC
mechanisms that play a role in electronuclear hadron production.  It
was shown that a reasonable description of the $^{16}$O(e,e$'$pp)
experimental data from NIKHEF could be obtained.  These two-nucleon
knockout data seem to provide some evidence for the unexpectedly
``low'' spectroscopic factors that were obtained from the analysis of
quasi-elastic (e,e$'$p) reactions.  Moreover, it turned out that the
SRC effects can be clearly separated from the $\Delta$-isobar
background at low missing ``diproton'' momenta in accordance with the
physical picture that the correlations in the single-particle spectral
function $P(\vec{k},E)$ are localized along a ridge imposed by the
kinematical constraints of heavy repulsion between the individual
nucleons that constitute pairs.  It was further shown that
polarization observables as they can be obtained from
($\overrightarrow{\mathrm{e}}$,e$'\overrightarrow{\mathrm{p}}$p)
measurements offer possibilities to further isolate the longitudinal
channel and to minimize at the same time the uncertainties with
respect to the final state interaction.  The usefulness of the
improved insight into the effect of two-nucleon degrees of freedom for
the electromagnetic probe was illustrated for double-polarization
single-nucleon knockout studies.

\section*{Acknowledgments}
This work is supported by the Fund for Scientific Research (FWO) -
Flanders under contract number G.0008.95.    
Stimulating discussions with G.~Rosner are gratefully acknowledged. 



\begin{thebibliography}{99}
%
\bibitem{vijay} 
V. Phandaripande, I. Sick and P.K.A. de Witt Huberts,
Rev. Mod. Phys. {\bf 69} (1997) 981.
\bibitem{ciofi}
C. Ciofi degli Atti and S. Simula, Phys. Rev. C {\bf 53} (1996) 1689.
\bibitem{strikman} L.L. Frankfurt and M.I. Strikman, Phys. Rep. {\bf
160} (1988) 236.   
\bibitem{jannpa1}
J. Ryckebusch, M. Vanderhaeghen, L. Machenil and M. Waroquier,
Nucl. Phys. {\bf A568} (1994) 828. 
\bibitem{janeepp}
J.Ryckebusch, V. Van der Sluys, K. Heyde, H. Holvoet, W. Van Nespen, 
M. Waroquier and M. Vanderhaeghen
Nucl. Phys. {\bf A624} (1997) 581.
\bibitem{janprc} 
J. Ryckebusch, D. Debruyne and W. Van Nespen,
Phys. Rev. C {\bf 57} (1998) 1319.
\bibitem{veerlelarge} 
V. Van der Sluys, J. Ryckebusch and M. Waroquier,
Phys. Rev. C {\bf 55} (1997) 1982.
\bibitem{giam} Giampaolo Co$'$ and Antonio M. Lallena, Phys. Rev. C
{\bf 57} (1998) 145.
\bibitem{gerco1}
C.J.G. Onderwater {\em et al.}, Phys. Rev. Lett. {\bf 78} (1997) 4893.
\bibitem{gercoprl2} 
C.J.G. Onderwater {\em et al.}, Phys. Rev. Lett. {\bf 81} (1998) 2213.
\bibitem{louk} L. Lapik\`{a}s, contribution to this proceedings.
\bibitem{guenther1} G. Rosner, contribution to this proceedings.
\bibitem{dhe3} G. Kaschl, G. Mairle, H. Mackh, D. Hartwig and
U. Schwinn, Nucl. Phys. {\bf A178} (1971) 275. 
\bibitem{cohen}  S. Cohen and D. Kurath, Nucl. Phys. {\bf A141}
(1977) 145. 
\bibitem{geurts} W.J.W. Geurts, K. Allaart,
W.H. Dickhoff, H. M\"uther, Phys. Rev. C {\bf 54} (1996) 1144.
\bibitem{carlottub} G. Giusti, F.D. Pacati, K. Allaart, W. Geurts,
H. Muether and W. Dickhoff, Phys. Rev. C {\bf 57} (1998) 1691.
\bibitem{sandorfi} A.M. Sandorfi and W. Leidemann, Phys. Rev. C {\bf
53} (1996) 1506. 
\bibitem{hirenzaki} S. Hirenzaki, J. Nieves, E. Oset and
M.J. Vicente-Vacas, Phys. Lett. {\bf B304} (1993) 198.
\bibitem{gearhart} C.C.Gearhart, PhD thesis, Washington University
(St. Louis, 1994), unpublished and W. Dickhoff, private communication.
\bibitem{flee}
F.X. Lee, C. Bennhold and L.E. Wright, 
Phys. Rev. C {\bf 55} (1997) 318.
\bibitem{carlotearly}
C. Giusti and F.D. Pacati, Nucl. Phys. {\bf
A535} (1991) 573.
\bibitem{a197} MAMI experiment A1/1-97 ``Investigation of Short-Range
nucleon-nucleon correlations using the reaction
$^{16}$O(e,e$'$pp)$^{14}$C in superparallel kinematics'' (Contact
Persons J.~Friedrich and G.~Rosner)
\bibitem{sick} TJNAF experiment E-97-006 ``Correlated Spectral
Function and the (e,e$'$p) reaction mechanism'' (Spokesperson I.~Sick)
\bibitem{lu98} D.H. Lu, A.W. Thomas, K. Tsushima, A.G. Williams and
K. Saito, Phys. Lett.  {\bf B417} (1998) 217.
\bibitem{ron} R.D. Ransome, contribution to this proceedings.
\bibitem{kelly} J.J. Kelly, Phys. Rev. C {\bf 56} (1997) 2672.
\bibitem{veerlemec} V. Van der Sluys, J. Ryckebusch and M. Waroquier,
Phys. Rev. {\bf C49} (1994) 2695.

%




\end{thebibliography}
\end{document}